\documentclass[orivec,envcountsame]{llncs}

\usepackage{amsmath}
\usepackage{amssymb}
\usepackage{stmaryrd}
\usepackage{theorem}
\usepackage{bussproofs}
\usepackage{url}
\usepackage{color}
\usepackage{paralist}
\usepackage{setspace}

\newcommand\sbr[1]{\llbracket #1 \rrbracket}
\newcommand\sysc[1]{\langle\!\langle #1 \rangle\!\rangle}

\allowdisplaybreaks[1]

\newcommand\dom{\mathsf{dom}}

\newcommand\pp{\mathcal{P}}
\renewcommand\ss{\mathcal{S}}
\newcommand\sem[1]{\llbracket #1\rrbracket}
\newcommand\call[1]{\mathtt{call}\,#1}
\newcommand\ret[1]{\mathtt{ret}\,#1}
\newcommand\redd[2]{\xrightarrow[{#2}]{#1}}
\newcommand\rede{\longrightarrow}

\definecolor{lgrey}{rgb}{0.5,0.5,0.5}
\definecolor{dblue}{rgb}{0,0,.75}

\newcommand{\ntnote}[1]{}
\newcommand\comp[2]{\otimes_{#1}^{#2}}
\newcommand\pr{\mathit{Pr}}
\newcommand\pub{\varPi}
\newcommand\prc{K}
\newcommand\cont[1]{\kappa(#1)}
\newcommand\func[1]{\phi(#1)}
\newcommand\loc[1]{\lambda(#1)}
\newcommand\clos[2]{Cl(#1,#2)}

\newcommand\proc[1]{\langle #1 \rangle}
\newcommand\NAM{\mathcal N}
\newcommand\NAMl{\NAM_{\lambda}}
\newcommand\NAMf{\NAM_{\phi}}
\newcommand\NAMc{\NAM_{\kappa}}

\newcommand\XX{\mathcal{X}}

\newcommand\Z{\mathbb{Z}}

\begin{document}
\title{A System-Level Game Semantics}
\author{Dan R. Ghica\inst{1} \and Nikos Tzevelekos\inst{2}}
\institute{University of Birmingham \and Queen Mary, University of London}


\maketitle

\begin{abstract}
  Game semantics is a trace-like denotational semantics for programming languages where the notion of legal observable behaviour
  of a term is defined combinatorially, by means of rules of a game between the term (the \emph{Proponent}) and its context (the \emph{Opponent}). 
  In general, the richer the computational features a language has the less
  constrained the rules of the semantic game. In this paper we
  consider the consequences of taking this relaxation of rules to the
  limit, by granting the Opponent \emph{omnipotence}, that is,
  permission to play any move without combinatorial restrictions. However, we impose an epistemic restriction by
  not granting Opponent \emph{omniscience}, so that {Proponent} can have undisclosed secret moves. 
  We introduce a basic C-like programming language and we define such a semantic model for it. 
  We argue that the resulting semantics is an appealingly simple combination of operational and game semantics and we show how
  certain traces explain system-level attacks, i.e.~plausible attacks that are  realisable outside of the programming
  language itself. We also show how allowing {Proponent} to have secrets ensures that some desirable equivalences in the programming
  language are preserved.
\end{abstract}

\section{Introduction}
Game semantics came to prominence by solving the long-standing open
problem of \emph{full abstraction} for
PCF~\cite{AbramskyJM00,HylandO00} and it consolidated its status as a
successful approach to modelling programming languages by being used
in the definition of numerous other fully abstract programming
language models.
The approach of game semantics is to model computation as a formal 
interaction, called a \emph{game}, between a term and its context. 
Thus, a semantic game features two players: a \emph{Proponent}, representing the term, and an \emph{Opponent} (O), representing the context. 
The interaction is formally described by sequences of game moves, called \emph{plays}, and a term is modelled by a corresponding \emph{strategy}, that is, the set of all its possible plays.
To define a game semantics one needs to define what are the {rules} of the game and what are the abilities of the players.

For PCF games, the rules are particularly neat, corresponding to the so-called ``principles of polite conversation'': 
moves are divided into \emph{questions} and \emph{answers}; players must take turns; no
question can be asked unless it is made possible (\emph{enabled}) by an earlier relevant question; no answer can be given unless it is to the
most recent unanswered question. The legality constraints for plays
can be imposed as combinatorial conditions on sequences of moves.

Strategies also have combinatorial conditions which characterise the players rather than the game. They are uniformity conditions which
stipulate that if in certain plays P makes a certain move, in other plays it will make an analogous move. The simplest condition is
\emph{determinism}, which stipulates that in any strategy if two plays are equal up to a certain P move, their subsequent P moves must also be the same.
Relaxing some of the combinatorial constraints on plays and strategies
elegantly leads from models of PCF to models of more expressive programming
languages. For example, relaxing a condition called \emph{innocence}
leads to models of programming language with
state~\cite{AbramskyM96}, relaxing \emph{bracketing} leads to models
of programming languages with control~\cite{Laird97}, and in the absence
of \emph{alternation} we obtain languages for
concurrency~\cite{GhicaM08}.

\paragraph{Contribution.}
In this paper we consider the natural question of what happens if in a
game semantics we remove combinatorial constraints from O's behaviour.
Unlike conventional game models, our construction is asymmetric:
P behaves in a way determined by the programming language and its inherent limitations,
whereas O may represent plausible behaviour which is, however, not syntactically realizable neither in the 
language nor in some in obvious extensions. In this paper we will see that
such a model is, in a technical sense, well formed and that the notion of
equivalence it induces is interesting and useful.

We study such a relaxed game model using an idealized type-free C-like
language. The notion of \emph{available move} is modelled using a notion
of \emph{secret} similar to that used in models of security
protocols, formally represented using
\emph{names}. This leads to a notion of Opponent which is omnipotent
but not omniscient: it can make any available move in any order, but
some moves can be hidden from it. This is akin to Dolev-Yao
attacker model of security.

We show how {inequivalences} in this semantic model
 capture \emph{system-level attacks}, i.e.~behaviours of the
ambient system which, although not realizable in the language itself,
can be nevertheless enacted by a powerful enough system. Despite the
existence of such a powerful ambient system we note that many
interesting equivalences still hold. This provides evidence  
that questions of semantic equivalence can be formulated outside
the conventional framework of a \emph{syntactic} context.

Technically, the model is expressed in an operationalised version of
game semantics like Laird's~\cite{Laird07} and names are handled
 using  nominal sets~\cite{GabbayP99}.

\section{A system-level semantics}

\subsection{Syntax and operational semantics}

We introduce a simple untyped C-like language which is just expressive
enough to illustrate the basic concepts. A \emph{program} is a list of
\emph{modules}, corresponding roughly to files in C. A
\emph{module} is a list of function or variable declarations. An
exported variable or function name is globally visible, otherwise its
scope is the module. In extended BNF-like notation we write:
\begin{align*}
  & \mathit{Prog} ::= \mathit{Mod}^* \quad
  \mathit{Mod} ::= \mathit{Hdr}\;\mathit{Mod}\quad
  \mathit{Hdr} ::= \mathtt{export}\;\overline x;\mathtt{import}\;\overline x;\\
  &\mathit{Dcl} ::= \mathtt{decl}\; x = n;\,
  \mathit{Dcl}\,\mid\, \mathtt{decl}\;
  \mathit{Func};\,\mathit{Dcl}\,\mid\, \epsilon
\end{align*}
The header \textit{Hdr} is a list of names exported and imported by
the program, with \emph{x} an identifier (or list of identifiers $\overline x$) taken from an infinite set
$\mathcal N$ of \emph{names}, and $n\in\Z$.

As in
C, functions are available only in global scope and in uncurried form:
\begin{equation*}
  \emph{Func} ::= x(\overline x) \{ \mathtt{local}\;\overline x;\;
  \emph{Stm}\; \mathtt{return}\;\emph{Exp};\}
\end{equation*}
A function has a name and a list of arguments. In the body of the
function we have a list of local variable declarations followed by a list
of statements terminated by a return statement. 
Statements and expressions are (with $n\in\Z$).
\begin{align*}
  &  \mathit{Stm} ::= \epsilon \mid \mathtt{if} (\mathit{Exp}) \mathtt{then} 
  \{\emph{Stm}\} \mathtt{else} \{\mathit{Stm}\}; \mathit{Stm}
  \;\mid \mathit{Exp}{=}\mathit{Exp}; \mathit{Stm}
   \mid \mathit{Exp}(\mathit{Exp}^*); \mathit{Stm} \\
  &  \mathit{Exp}  ::= \mathit{Exp} \star \mathit{Exp}
  \mid *\mathit{Exp}  \mid \mathit{Exp}(\mathit{Exp}^*)
  \mid \mathtt{new()} \mid n
\end{align*}
Statements are branching, assignment and function call. For simplicity, iteration is
not included as we allow recursive calls. Expressions are
arithmetic and logical operators, variable dereferencing
($*$), variable allocation and integer constants.  Function call can be
either an expression or a statement. Because the language is type-free
the distinction between statement and expression is arbitrary and only
used for convenience.

If $\mathtt{decl\ }f(\overline x)\{e\}$ is a declaration in module $M$
we define $f\,@\,M=e[\overline x]$, interpreted as ``the definition of
$f$ in $M$ is $e$, with arguments $\overline x$.''

A \emph{frame} is given by the grammar below, with $op\in\{{=}, {\star}, {;} \}$,
$op'\in\{{*}, {-} \}$.
\begin{align*}
  t::=
  \mathtt{if\,} (\square)\, \mathtt{then\,}
  \{e\}\, \mathtt{else\,} \{e\}\,
  \mid\, \square\; op\; e \,
  \mid\, v \;op\; \square \,\mid\, op'\; \square \,
  \mid\, \square\; e\,
  \mid\, v\;\square\,
  \mid\, (\square, e)\,
  \mid\, (v, \square)
\end{align*}
We denote the ``hole'' of the frame by $\square$. We denote by
$\mathcal Fs$ the set of lists of frames, the \emph{frame stacks}. By $v$
we denote \emph{values}, defined below. 

Our semantic setting is that of nominal sets~\cite{GabbayP99}, constructed over
the multi-sorted set of names $ \NAM = \NAMl \uplus \NAMf \uplus \NAMc
$ where each of the three components is a countably infinite set of
\emph{location names}, \emph{function names} and \emph{function
  continuation names} respectively.  We range over names by $a,b$,
etc. Specifically for function names we may use $f$, etc.; and for continuation names $k$, etc.
For each set of names $\XX$ we write $\loc\XX$, $\func\XX$ and
$\cont\XX$ for its restriction to location, function and continuation
names respectively.  We write $\nu(x)$ for the \emph{support} of $x$,
for any element $x$ of a nominal set $X$, i.e.\ all the free names
occurring in it.

A store is defined as a pair of partial functions with
finite domain:
\[
s \in\mathit{Sto} = (\NAM_\lambda \rightharpoonup_{\mathsf{fn}} (\Z+\NAM_\lambda+\NAM_\phi))
\times (\NAMc \rightharpoonup_{\mathsf{fn}} \mathcal{F}s\times\NAMc)
\]
The first component of the store assigns integer values (data), other
locations (pointers) or function names (pointer to functions) to
locations. The second stores \emph{continuations},  
used by the system to resume a suspended function call.

We write $\loc s$, $\cont s$ for the two projections of a store
$s$. By abuse of notation, we may write $s(a)$ instead of $\loc s(a)$
or $\cont s(a)$. Since names are sorted, this is unambiguous.
The support $\nu(s)$ of $s$ is the set of names appearing in its domain or value
set.
For all stores $s,s'$ and set of names $\mathcal X$, we use
the notations:
\begin{description}
\item[restrict-to:] only consider the subset of a store defined at a given set of names, $s \upharpoonright \mathcal X =\ \{ (a,y)\in
  s\ |\ a\in \mathcal X\}$
\item[restrict-from:] only consider the subset of a store that is not defined
 at a given set of names, $s\setminus \mathcal X =\ s\upharpoonright(\dom(s)\setminus \mathcal
  X)$
\item[update:] change the values in a store, $s[a\mapsto x]=\{(a,x)\}\cup(s\setminus\{a\})$ and, more generally, $s[s'] = s'\cup
  (s\setminus\dom(s'))$
\item[valid  extension:] $s\sqsubseteq s'$ if
$\dom(s) \subseteq \dom(s')$
\item[closure:]  $\clos{s}{\XX}$ is the least set of names containing
$\XX$ and all names reachable from $\XX$ through $s$ in a transitively
closed manner, i.e.\
$ \XX\subseteq \clos{s}{\XX}$ and if $(a,y)\in s$ with $a\in\clos{s}{\XX}$ then
$\nu(y)\in\clos{s}{\XX}.$
\end{description}
We give a semantics for the language using a frame-stack abstract
machine.
It is convenient to
take \emph{identifiers} to be
\emph{names},
as it gives a simple way to handle pointers to functions
in a way much like that of the C language.
We define a \emph{value}
to be a name, an integer, or a tuple of values:
$
v ::= ()\mid a \mid n \mid (v,v).
$
The value $()$ is the unit for the tuple operation. Tupling is
associative and for simplicity we identify tuples up to associative isomorphism,
so $(v,(v,v))=((v,v),v)=(v,v,v)$ and $(v,())=v$, etc.
If a term is not a value we write it as $e$.

\newcommand\f{t}
The \emph{Program configurations} of the abstract machine are
\[
 \langle N\mid
 P \vdash s, \f, e, k \rangle \in \NAM\times\NAM\times
 \mathit{Sto}\times \mathcal{F}s\times \mathit{Exp}\times\NAMc
\]
$N$ is a set of \emph{used names}; $P\subseteq N$ is the set of
\emph{public names}; $s$ is the \emph{program state}; $\f$ is a list of
frames called the \emph{frame stack}; $e$ is the (closed) expression,
 being evaluated; and $k$ is a \emph{continuation name}, which for
now will stay unchanged.

The transitions of the abstract machine
\[
\langle
N\mid P \vdash
s, \f, e, k
\rangle \longrightarrow
\langle
N'\mid P' \vdash
s', \f', e', k
\rangle
\]
are defined by case analysis on the structure of $e$ then $\f$ in a standard fashion,
as in Fig.~\ref{fig:os}.
\begin{figure}[t]\small
\paragraph{Case $e=v$ is a value.}
\begin{align*}
  &\langle N\mid P \vdash
  s, \f\circ(\mathtt{if\;} (\square)\; \mathtt{then\;}
  \{e_1\}\; \mathtt{else\;} \{e_2\}),v
  ,k\rangle
  \longrightarrow
  \langle N\mid P \vdash
  s, \f, e_1
  ,k\rangle, \text{ if }v\in\Z\setminus\{0\}
  \\&
  \langle N\mid P \vdash
  s,\f\circ(\mathtt{if\;} (\square)\; \mathtt{then\;}
  \{e_1\}\; \mathtt{else\;} \{e_2\}),v
  ,k\rangle
  \longrightarrow
  \langle N\mid P \vdash
  s, \f, e_2
  ,k\rangle, \text{ if }v= 0
  \\&
  \langle  N\mid P \vdash
  s, \f\circ(\square\;op\;e),v
  ,k\rangle \longrightarrow \langle N\mid P \vdash
  s,\f\circ(v\;op\;\square), e,k\rangle
  \text{ for }op\in\{{=}, {\star}, ; \}
  \\&
  \langle  N\mid P \vdash
  s,\f\circ(v\star\square), v',k\rangle
  \longrightarrow
  \langle N\mid P \vdash
  s,\f,v'',k\rangle,\text{ and }v''=v\star v'
  \\&
  \langle  N\mid P \vdash
  s,\f\circ(v;\square), v',k\rangle
  \longrightarrow
  \langle N\mid P \vdash
  s,\f,v',k\rangle
  \\&
  \langle  N\mid P \vdash
  s,\f\circ(a=\square), v,k\rangle
  \longrightarrow
  \langle N\mid P \vdash
  s[a\mapsto v],\f,(),k\rangle
  \\&
  \langle  N\mid P \vdash
  s,\f\circ(*\square), v,k\rangle
  \longrightarrow
  \langle N\mid P \vdash
  s,\f,s(v),k\rangle
  \\&
  \langle  N\mid P \vdash
  s,\f\circ(\square;e), \mathtt{local}\;x,k\rangle
  \longrightarrow
  \langle N\cup\{a\}\mid P \vdash
  s[a\mapsto 0], \f,e[a/x],k\rangle,\text{ if
  }a\not\in N
  \\&
  \langle  N\mid P \vdash
  s,\f\circ(\square(e)), v,k\rangle
  \longrightarrow
  \langle N\mid P \vdash
  s,\f\circ(v(\square)),e,k\rangle
  \\&
  \langle  N\mid P \vdash
  s,\f\circ((\square,e)),v,k\rangle
  \longrightarrow
  \langle  N\mid P \vdash
  s,\f\circ((v,\square)),e,k\rangle
  \\&
  \langle  N\mid P \vdash
  s,\f\circ((v,\square)),v',k\rangle
  \longrightarrow
  \langle  N\mid P \vdash
  s,\f,(v,v'),k\rangle
  \\&
  \langle  N\mid P \vdash
  s,\f\circ(f(\square)),v',k\rangle
  \longrightarrow
  \langle  N\mid P \vdash
  s,\f, e[v'/\overline x],k\rangle,\
  \text{if } f\,@\,M=e[\overline x] \tag{F}\label{FC}
 \end{align*}
 \paragraph{Case $e$ is not a canonical form.}
 \begin{align*}
   &\langle N\mid P \vdash s, \f,\mathtt{if\;} (e)\;
  \mathtt{then\;} \{e_1\}\; \mathtt{else\;} \{e_2\} ,k\rangle
  \longrightarrow
  \langle N\mid P \vdash s, \f\circ(\mathtt{if\;} (\square)\;
  \mathtt{then\;} \{e_1\}\; \mathtt{else\;} \{e_2\}),e ,k\rangle
  \\&
  \langle  N\mid P \vdash
  s, \f, e\;op\;e'
  ,k\rangle \longrightarrow \langle N\mid P \vdash
  s,\f\circ(\square\;op\;e'), e,k\rangle, \text{if } op\in\{{=},
  {\star}, ; \}
  \\&
  \langle  N\mid P \vdash
  s, \f, op\;e
  ,k\rangle\longrightarrow \langle N\mid P \vdash
  s,\f\circ(op\;\square), e,k\rangle,\text{if }op\in\{\mathtt{return}(-),{*}\}
  \\&
  \langle  N\mid P \vdash
  s, \f, \mathtt{new}()
  ,k\rangle\longrightarrow \langle N\cup\{a\}\mid P \vdash
  s[a\mapsto 0],\f, a,k\rangle, \text{if
  }a\in\NAM_\lambda\setminus N
  \\&
  \langle  N\mid P \vdash
  s, \f, e(e')
  ,k\rangle\longrightarrow \langle N\mid P \vdash
  s,\f\circ(\square(e')), e,k\rangle
  \\&
  \langle  N\mid P \vdash
  s, \f, (e,e')
  ,k\rangle\longrightarrow\langle N\mid P \vdash
  s,\f\circ((\square,e')), e,k\rangle
\end{align*}
\caption{Operational semantics}
\label{fig:os}
\end{figure}
Branching is as in C, identifying non-zero values with true and zero
with false. Binary operators are evaluated left-to-right, also as in
C.  Arithmetic and logic operators ($\star$) have the obvious
evaluation. 
Dereferencing
is given the usual evaluation, with a note that in order for the rule
to apply it is implied that $v$ is a location and $s(v)$ is
defined. Local-variable allocation extends the domain of $s$ with a
fresh secret name.  Local variables are created fresh, locally for the scope
of a function body. The \texttt{new}() operator allocates a secret and fresh location name, initialises it to zero and
returns its location. The return statement is used as a syntactic marker
for an end of function but it has no semantic role.

Structural rules, such as function application and tuples are as usual
in call-by-value languages, i.e.\ left-to-right.  Function call also
has a standard evaluation. The body of the function replaces the
function call and its formal arguments $\overline x$ are substituted
by the tuple of arguments $v'$ in point-wise fashion. Finally,
non-canonical forms also have standard left-to-right evaluations.

\subsection{System semantics}

The conventional function-call rule~\eqref{FC}
is only applicable if there is a function definition in the module. If the
name used for the call is not the name of a known function then the
normal operational semantics rules no longer apply.
We let function calls and returns, when the function is not locally
defined, be a mechanism for interaction between the program and the
ambient system.  A \emph{System configuration} is a triple $
\sysc{N\mid P\vdash s} \in \NAM\times\NAM\times \mathit{Sto}.  $

Given a module $M$ we will write as $\sbr{M}$ a transition system
defining its system-level semantics (SLS). Its states are $\mathcal S
\sbr M = Sys\sbr M\cup Prog\sbr M $, where $Prog\sbr M$ is the set of
abstract-machine configurations of the previous section and $Sys\sbr
M$ is the set of system configurations defined above.
The SLS is defined at the level of modules, that is programs with
missing functions, similarly to what is usually deemed a
\emph{compilation unit} in most programming languages.

Let $\mathcal L_{PS}\simeq\mathcal L_{SP}=\{\mathtt{call}\,f,v,k\mid
f\in\NAM_\lambda,\,k\in\NAM_\kappa,\,v\text{ a value}\}\cup \{\mathtt{ret}\,v,k\mid k\in\NAM_\kappa,\,v\text{
 a value}\}$.
The transition relation is of the form:

{\allowdisplaybreaks[0]
\begin{multline*}
  \delta \sbr {M} \subseteq  (Prog\sbr M\times Prog\sbr
  M) \cup (Prog\sbr M\times \mathcal
  L_{PS}\times\mathit{Sto}\times Sys\sbr M)\\ \cup (Sys\sbr
  M\times \mathcal L_{SP}\times\mathit{Sto}\times Prog\sbr M)
\end{multline*}
}

In transferring control between Program and System the continuation
pointers ensure that upon return the right
execution context can be recovered. We impose several hygiene
conditions on how continuations are used, as follows.
We distinguish between P-continuation names and S-continuation
names. The former are created by the Program and stored for subsequent
use, when a function returns. The latter are created by the System and
are not stored. The reason for this distinction is both technical and
intuitive. Technically it will simplify proving that composition
is well-defined. 
%
Intuitively, mixing S and P continuations does not
create any interesting behaviour. If S gives P a continuation it does
not know then P can only crash. It is not interesting for S to make P
crash, because S can crash directly if so it chooses. So this is
only meant to remove trivial behaviour.

The first new rule, called program-to-system call is:
\\[1.5ex]
\framebox{
  \begin{minipage}[t]{\linewidth}
    \paragraph{Program-to-System call:}
    \[
      \langle N\mid P \vdash s,\f\circ(f(\square)),v,k\rangle
      \xrightarrow[{s\upharpoonright
        \loc{P'}}]{\mathtt{call}\,f,v,k'} \sysc{N\cup\{k'\}\mid
        P'\cup\{k'\} \vdash s[k'\mapsto (\f,k)]}
    \]
    if $f\,@\,M$ not defined, $k'\notin N,P'=Cl(s,P\cup\nu(v))$.
  \end{minipage}
} 
\\[1.5ex]
When a non-local function is called, control is transferred to the
system. In game semantics this 
corresponds to a \emph{Proponent question}, and is an observable action. Following
it, all the names that can be transitively reached from public
names in the store also become public, so it gives both control and information to the System.
Its
observability is marked by a label on the transition arrow, which
includes: a tag \texttt{call}, indicating that a function is called,
the name of the function ($f$), its arguments ($v$) and a
fresh continuation ($k'$), which stores the
code pointer; the transition also marks that part of the store which
is observable because it uses publicly known names.

The counterpart rule is the system-to-program return,
corresponding to a return from a non-local function.
\\[1.5ex]
\fbox{
\begin{minipage}{\linewidth}
  \paragraph{System-to-Program return:}
  \[
    \sysc{N\mid P\vdash s}\xrightarrow[s']{\mathtt{ret}\,v,k'}
    \langle N\cup\nu(v,s') \mid P\cup\nu(v,s')\vdash s[s'],f,v,k\rangle 
  \]
  if $s(k')=(f,k), \nu(v,s')\cap N\subseteq P,
  \loc{\nu(v)}\subseteq\nu(s'), s\upharpoonright
    \loc P\sqsubseteq s', \cont{s'}=\emptyset$.
\end{minipage}
}\\[1.5ex]
This is akin to the game-semantic  \emph{Opponent answer}. Operationally it
corresponds to S returning from a function. Note here that the only
constraints on what S can do in this situation are \emph{epistemic,}
i.e.\ determined by what it \emph{knows}:
\begin{enumerate}
\item it can return with any value $v$ so long as it only
  contains public names or fresh names (but not \emph{private} ones);
\item it can update any public location with any value;
\item it can return to any (public) continuation $k'$.
\end{enumerate}
However, the part of the store which is private (i.e.\ with domain in $N\setminus P$) cannot be modified by
S. So S has no restrictions over what it can do with known names and
to known names, but it cannot guess private names. Therefore it cannot
do anything with or to names it does not know. The restriction on the
continuation are just hygienic, as explained earlier.

There are two converse transfer rules corresponding to the program
returning and the system initiating a function call:
\\[1.5ex]
\fbox{
\begin{minipage}{\linewidth}
  \paragraph{System-to-Program call:}
  \[
    \sysc {N\mid P\vdash s} \xrightarrow[s']{\mathtt{call}\,f,v,k} \langle N\cup\{k\}\cup \nu(v,s') \mid P\cup\{k\}\cup
    \nu(v,s')\vdash s[s'],f(\square),v,k\rangle 
  \]
  if $f\,@\,M$ defined, $k\notin\dom(s),\nu(f,v,s')\cap N\subseteq P,\loc{\nu(v)}\subseteq\nu(s'),s\upharpoonright \loc P\sqsubseteq s',\cont {s'}=\emptyset$.
\\[-1ex]
  \paragraph{Program-to-System return:}
    \[
    \langle N\mid P\vdash s,-,v,k \rangle
    \xrightarrow[s\upharpoonright \loc {P'}]{\mathtt{ret}\,
      v,k}
    \sysc {N\mid P'\vdash s},
    \text{where }P'=Cl(s,P\cup\nu(v)).
	\]
\end{minipage}
}\\[1.5ex]
In the case of the S-P call it is S which calls a publicly-named
function from the module. As in the case of the return, the only
constraint is that the function $f$, arguments $v$ and the state
update $s'$ only involve public or fresh names. The hygiene conditions on the
continuations impose that no continuation names are stored, for reasons already explained.
Finally, the P-S return represents the action of the program yielding
a final result to the system following a function call. The names used
in constructing the return value are disclosed and the public part of
the store is observed.
In analogy with game semantics the function return is a
\emph{Proponent answer} while the system call is an \emph{Opponent
  question}.

The \emph{initial configuration} of the SLS for module $M$ is $S^0_M=\sysc {N\mid P\vdash s_0}$. It contains a store
$s_0$ where all variables are initialised to the value specified in the declaration. The set $N$ of names
contains all the exported and imported names, all declared variables
and functions. The set $P$ contains all exported and imported names.

\section{Compositionality}\label{sec:comp}

The SLS of a module $M$
gives us an interpretation $\sbr M$ which is modular and effective
(i.e.\ it can be executed) so no consideration of the context is
required in formulating properties of modules based on their
SLS. Technically, we can reason about SLS using standard tools for
transition systems such as trace equivalence, bisimulation or
Hennessy-Milner logic.

We first show that the SLS is
consistent by proving a \emph{compositionality} property. SLS
interpretations of modules can be composed semantically in a way that
is consistent with syntactic composition. Syntactic composition for
modules is concatenation with renaming of un-exported function and
variable names to prevent clashes, which we will denote by using
$-\cdot-$.  We call this \emph{the principle of functional
  composition}.

In this section we show that we can define a semantic SLS
composition {$\comp{}{}$} so that, {for an appropriate notion of bisimulation in
  the presence of bound names and $\tau$-transitions:}
\\[1.5ex]\fbox{
  \begin{minipage}{0.95\linewidth}
    \paragraph{Functional composition.} For any modules $M,M'$:
    $
    {    \sbr{M\cdot M'}\sim\sbr{M}\comp{}{}\sbr{M'}.}
    $
  \end{minipage}
}\\[0ex]

Let $\pp$ range over program configurations, and $\ss$ over system
configurations. We define semantic composition of modules inductively as in Fig.~\ref{fig:semcom} (all rules have symmetric,
omitted counterparts).  {We use an extra component $\pub$ containing
  those names which have been communicated between either module and the outside
  system, and we use an auxiliary store $s$ containing values of
  locations only.
{Continuation names in each $\pub$ are assigned Program/System polarities, thus specifying whether a continuation name was introduced by either of the modules or from the outside system. We write $k\in\pub_P$ when $k\in\pub$ has Program polarity, and dually for $k\in\pub_S$.}
We also use the following notations for
  updates, where we write $\pr$ for the
set of private names $\nu(\ss,\ss')\setminus\pub$.
\begin{align*}
&(\pub,s')^P[v,k,s] = (\pub',s'[s])\text{ where }\pub'=\clos{s'[s]}{\nu(v)\cup\pub}\cup\{k\}\text{ and }k\in\pub'_P\\
&(\pub,s')^S[v,k,s] = (\pub',s'[s])\text{ where }\pub'=\pub\cup\nu(v,s\setminus\pr)\cup\{k\}\text{ and }k\in\pub'_S\\
&\qquad\qquad\qquad\qquad\qquad\qquad\text{ and } (s'\upharpoonright\pub)\sqsubseteq s,\,s'\setminus\pub\subseteq s,\,
        \nu(v',s\setminus\pr)\cap\pr=\emptyset
\end{align*} 
The same notations are used when no continuation name $k$ is included in the update.
Private calls and returns are assigned $\tau$-labels, thus specifying the fact that they are internal transitions.}
\begin{figure*}[t]
  \begin{enumerate}
    \item \label{mv:int}Internal move:
    \AxiomC{$\pp\rede \pp'$}
    \RightLabel{$\nu(\pp')\cap\nu(\ss)\subseteq\nu(\pp)$}
    \UnaryInfC{$\pp\comp{\pub}{s}\ss\rede \pp'\comp{\pub}{s}\ss$}
    \DisplayProof
    \item \label{mv:cal} Cross-call:
    \AxiomC{$\pp\redd{\call{f,v,k}}{s}\ss' $}
    \AxiomC{$\ss\redd{\call{f,v,k}}{s}\pp' $}
    \RightLabel{$k\notin\nu(\ss)$}
    \BinaryInfC{$\pp\comp{\pub}{s'}\ss\redd{\tau}{}
      \ss'\comp{\pub}{s'[s]}\pp' $} \DisplayProof
    \item \label{mv:ret} Cross-return:
    \AxiomC{$\pp\redd{\ret{v,k}}{s}\ss' $}
    \AxiomC{$\ss\redd{\ret{v,k}}{s}\pp' $}
    \BinaryInfC{$\pp\comp{\pub}{s'}\ss\redd{\tau}{}
      \ss'\comp{\pub}{s'[s]}\pp' $} \DisplayProof 
    \item \label{mv:progc} Program call:
    \AxiomC{$\pp\redd{\call f,v,k}{s}\ss' $}
    \AxiomC{$\ss\;\;\;\;\;\;\not\!\!\!\!\!\!\!\!\!\!\redd{\call
        f,v,k}{s}\pp' $}
    \RightLabel{$\begin{aligned}&(\pub',s'')=(\pub,s')^P[v,k,s]\\[-1mm]
    &k\notin\nu(\ss)\end{aligned}$}
    \BinaryInfC{$\pp\comp{\pub}{s'}\ss\redd{\call
        f,v,k}{s''\upharpoonright\pub'} \ss'\comp{\pub'}{s''}\ss $}
    \DisplayProof
    \item \label{mv:progr} Program return:
    \AxiomC{$\pp\redd{\ret v,k}{s}\ss' $}
    \AxiomC{$\ss\;\;\;\;\not\!\!\!\!\!\!\redd{\ret v,k}{s}\pp' $}
    \RightLabel{$(\pub',s'')=(\pub,s')^P[v,s]$}
    \BinaryInfC{$\pp\comp{\pub}{s'}\ss\redd{\ret
        v,k}{s''\upharpoonright\pub'} \ss'\comp{\pub'}{s''}\ss $}
    \DisplayProof
    \item \label{mv:sysc} System call:
    \AxiomC{$\ss\redd{\call f,v,k}{s}\pp $}
    \RightLabel{$\begin{aligned}&(\pub',s'')=(\pub,s')^S[v,k,s]\\[-1mm] 
    &k\notin\nu(\ss')\setminus\pub_S
      \end{aligned}$} \UnaryInfC{$\ss\comp{\pub}{s'}\ss'\redd{\call
        f,v,k}{s\upharpoonright\pub'} \pp\comp{\pub'}{s''}\ss'$}
    \DisplayProof
    \item \label{mv:sysr} System return:
    \AxiomC{$\ss\redd{\ret v,k}{s}\pp $} \RightLabel{$\begin{aligned}
        &(\pub',s'')=(\pub,s')^S[v,s]\\[-1mm] & k\notin\nu(\ss')\end{aligned}$}
    \UnaryInfC{$\ss\comp{\pub}{s'}\ss'\redd{\ret
        v,k}{s\upharpoonright\pub'} \pp\comp{\pub'}{s''}\ss'$}
    \DisplayProof
  \end{enumerate}

 \caption{Rules for semantic composition}
\label{fig:semcom}
\end{figure*}

The semantic composition of modules
${M}$ and ${M'}$ is given by
$
\sem{M}\comp{}{}\sem{M'} = \sem{M}\comp{\pub_0}{s_0\cup s_0'}\sem{M'},
$
where $s_0$ is the store assigning initial values to all initial public locations of $\sem{M}$, and
similarly for $s_0'$, and $\pub_0$ contains all exported and imported names.
The rules of Fig.~\ref{fig:semcom} feature side-conditions regarding
name-privacy. These stem from nominal game semantics~\cite{Laird04}
and they guarantee that the names introduced (freshly) by $M$ and $M'$
do not overlap and that the names introduced by the system in the
composite module do not overlap with any of the names introduced by
$M$ or $M'$. The former condition is necessary for correctness. The
latter is typically needed for associativity.

Let us call the four participants in the composite SLS \emph{Program
  A}, \emph{System A, Program B, System B}. Whenever we use X, Y as
Program or System names they can be either A or B, but
different. Whenever we say \emph{Agent} we mean Program or
System. System X and Program X form \emph{Entity X}. A state of the
composite system is a pair (Agent X, Agent Y) noting that they cannot be
both Programs. The composite transition rules reflect the following
intuitions:
\begin{itemize}
\item Rule~\ref{mv:int}: If Program X makes an internal (operational) transition System Y
  is not affected.
\item Rules~\ref{mv:cal}-\ref{mv:ret}: If Program X makes a system
  transition to System X and System Y can match the transition going
  to Program Y then the composite system makes an internal transition.
  This is the most important rule and it is akin to game semantic
  composition via ``synchronisation and hiding''. It signifies $M$
  making a call (or return) to (from) a function present in $M'$.
\item Rules~\ref{mv:progc}-\ref{mv:progr}: If Program X makes a system
  transition that cannot be matched by Entity Y then it is a system
  transition in the composite system, a non-local call or return.
\item Rules~\ref{mv:sysc}-\ref{mv:sysr}: From a composite system
  configuration (both entities are in a system configuration) either
  Program X or Program Y can become active via a call or return from
  the system.
\end{itemize}

\begin{lemma}
Let $X_1\comp{\pub}{s}X_2$ be a state in the transition graph of $\sem{M}\comp{}{}\sem{M'}$ that is reachable from the initial state. Then, if each $X_i$ includes the triple $(N_i,P_i,s_i)$, the following conditions hold.
\ntnote{These should follow from the definitions.}
\begin{itemize}
\item $(N_1\setminus P_1)\cap N_2 = N_1\cap(N_2\setminus P_2)=\emptyset$, $P_1\setminus\pub=P_2\setminus\pub$, $\pub\subseteq\nu(s)\cup\cont{P_1\cup P_2}\subseteq P_1\cup P_2$ and $\nu(\dom(s))=\loc{P_1\cup P_2}$.
\item $\dom(\cont{s_1})\cap\dom(\cont{s_2})=\emptyset$, $(\dom(\cont{s_1})\cup\dom(\cont{s_2}))\cap\pub_S=\emptyset$ and $\cont{P_1\cap P_2}\setminus\pub_S=\cont{P_1\cup P_2}\setminus\pub$.
\item If both $X_1,X_2$ are system configurations and $X_1\comp{\pub}{s}X_2$ is preceded by a state of the form $\pp\comp{\pub'}{s'}\ss$ then $s\upharpoonright P_1\subseteq s_1$ and $s\upharpoonright(P_2\setminus P_1)\subseteq s_2$, and dually if preceded by $\ss\comp{\pub'}{s'}\pp$. Thus, in both cases, $s\upharpoonright(P_i\setminus(P_1\cap P_2))\subseteq s_i$ for $i=1,2$.
\item Not both $X_1,X_2$ are program configurations. If $X_i$ is a program configuration then $s\upharpoonright(P_{3-i}\setminus P_i)\subseteq s_{3-i}$.
\end{itemize}
\end{lemma}

Semantic composition introduces a notion of private names: internal continuation names passed around between the two modules in order to synchronise their mutual function calls. As the previous lemma shows, these names remain private throughout the computation. Therefore, in checking bisimilarity for such reduction systems, special care has to be taken for these private names so that external system transitions capturing them do not affect these checks. This is standard procedure in calculi with name-binding.

We define the following translation $R$ from reachable composite states of $\sem{M}\comp{}{}\sem{M'}$ to states of $\sem{M\cdot M'}$.
\begin{align*}
&\sysc{N_1\mid P_1\vdash s_1}\comp{\pub}{s}\sysc{N_2\mid P_2\vdash s_2}
\longmapsto \sysc{(N_1\cup N_2)\setminus \prc\mid \pub\vdash (\hat s_1[s']\cup \hat s_2[s'])\setminus \prc}\\[-.2mm]
&\sysc{N_1\mid P_1\vdash s_1}\comp{\pub}{s}\proc{N_2\mid P_2\vdash s_2,\f,v,k}
\longmapsto \proc{(N_1\cup N_2)\setminus \prc\mid \pub \vdash \hat s_1[\hat s_2]\setminus \prc,\f',v,k'}\\[-.2mm]
&\proc{N_1\mid P_1\vdash s_1,\f,v,k}\comp{\pub}{s}\sysc{N_2\mid P_2\vdash s_2}\longmapsto \proc{(N_1\cup N_2)\setminus \prc\mid \pub\vdash \hat s_2[\hat s_1]\setminus\prc,\f',v,k'}\\[-6mm]
\end{align*}
where $K=\cont{P_1\cap P_2}\setminus\pub_S$, $s'=s\upharpoonright(P_1\cap P_2)$, $\hat s_i=s_i[k\mapsto (s_1,s_2)_\prc(s_i(n))]$ for all $k\in\dom(\cont{s_i})$, and $(\f',k')=(s_1,s_2)_K(\f,k)$. The function $(s_1,s_2)_\prc$ fetches the full external frame stack and the external continuation searching back from $(\f,k)$, that is,
$(s_1,s_2)_\prc(\f,k) = (\f,k)$ if $k\notin\prc$, otherwise if $k\in K$ and $s_i(k)=(\f',k')$ then $(s_1,s_2)_\prc(\f'\circ \f,k')$.

The translation merges names from the component configurations and deletes the names in $\prc$: these private names do not appear in $\sem{M\cdot M'}$, as there the corresponding function calls happen without using the call-return mechanism.
It also sets $\pub$ as the set of public names.
Moreover, the total store is computed as follows. In system configurations we just take the union of the component stores and update them with the values of $s$, which contains the current values of all common public names. In program configurations we use the fact that the P-component contains more recent values than those of the S-component.
\begin{proposition}\label{prop:compsls}
For $R$ defined as above and $X_1\comp{\pub}{s}X_2$ a reachable configuration,
\begin{enumerate}
\item if $X_1{\comp{\pub}{s}}X_2\redd{\tau}{} X_1'{\comp{\pub}{s'}}X_2'$ \hbox{then $R(X_1{\comp{\pub}{s}}X_2)=R(X_1'{\comp{\pub}{s'}}X_2')$,}
\item if $X_1{\comp{\pub}{s}}X_2\rightarrow X_1'{\comp{\pub}{s}}X_2'$ then $R(X_1{\comp{\pub}{s}}X_2)\rightarrow R(X_1'{\comp{\pub}{s}}X_2')$,
\item if $R(X_1\comp{\pub}{s}X_2)\redd{}{} Y$ and $X_1\comp{\pub}{s}X_2\not\redd{\tau}{}$ \ \ then $X_1\comp{\pub}{s}X_2\redd{}{} X_1'\comp{\pub}{s}X_2'$ with $Y=R(X_1'\comp{\pub}{s}X_2')$,
\item if $X_1\comp{\pub}{s}X_2\redd{\alpha}{s'}X_1'\comp{\pub'}{s''}X_2'$ then $R(X_1\comp{\pub}{s}X_2)\redd{\alpha}{s'}{R(X_1'\comp{\pub'}{s''}X_2')}$,
\item if $R(X_1\comp{\pub}{s}X_2)\redd{\alpha}{s'} Y$, $X_1\comp{\pub}{s}X_2\not\redd{\tau}{}$ \ \ and $\nu(\alpha)\cap K=\emptyset$ then $X_1\comp{\pub}{s}X_2\redd{\alpha}{s'} X_1'\comp{\pub'}{s''}X_2'$ with $Y={R(X_1'\comp{\pub'}{s''}X_2')}$,
\end{enumerate}
where $K$ is obtained from $X_1\comp{\pub}{s}X_2$ as above.
\end{proposition}
With bisimilarity between semantic and syntactic composites defined as above, functional composition follows immediately as a consequence.

\section{Reasoning about SLS}

The epistemically-constrained system-level semantics gives a security-flavoured
semantics for the programming language which is reflected by its logical 
properties and by the notion of equivalence it gives rise to.

We will see that certain properties of traces in the SLS of a module correspond
to ``secrecy violations'', i.e.\ undesirable disclosures of names that are meant
to stay secret. In such traces it is reasonable to refer to the System as an 
\emph{attacker} and consider its actions an attack. We will see that although
the attack cannot be realised within the given language it can be enacted
in a realistic system by system-level actions. 

We will also see that certain equivalences that are known to hold in conventional semantics still
hold in a system-level model. This means that even in the presence of an omnipotent attacker,
unconstrained by a prescribed set of language constructs, the
epistemic restrictions can prevent certain observations, not only by
the programming context but by any ambient computational system. This is a very powerful
notion of equivalence which embodies \emph{tamper-resistance} for  a module.

\paragraph{Note.} We chose these examples to illustrate the conceptual
interest of the SLS-induced properties rather than as an
illustration of the mathematical power of SLS-based reasoning
techniques. For this reason, we chose examples which are as simple and
clear as possible.

\subsection{A system-level attack: violating secrecy}\label{sec:vio}
This example is inspired by a flawed security protocol which is
informally described as follows.
\begin{center}
  \parbox{.95\linewidth}{Consider a secret, a locally generated key and an item of data read
  from the environment. If the local key and the input data are equal
  then output the secret, otherwise output the local key.}
\end{center}
In a conventional process-calculus syntax the protocol can be written as
\[
\nu s\nu k.\mathsf{in}(a).\mathsf{if}\, k{=}a\, \mathsf{then}\,
\mathsf{out}(s)\, \mathsf{else}\, \mathsf{out}(k).\]
It is true that
the secret $s$ is not leaked because the local $k$ cannot be known as
it is disclosed only at the very end. This can be proved using
bisimulation-based techniques for anonymity.
Let us consider an implementation of the protocol:
\begin{verbatim}
    export prot;
    import read;
    decl prot( ) {  local s, k, x;
                    s = new(); k = new(); x = read();
                    if (*x == *k) then *s else *k  }
\end{verbatim}
We have local variables \texttt{s} holding the ``secret location''
and \texttt{k} holding the ``private location''. We use the non-local,
 system-provided, function \texttt{read} to obtain a name from the
system, which cannot be that stored at \texttt{s} or \texttt{k}. A
value is read into \texttt{x} using untrusted system call
\texttt{read()}. Can the secrecy of \texttt{s} be violated by making
the name stored into it public? Unlike in the process-calculus model,
the answer is ``yes''.

\newcommand{\qwes}{\mathtt{s}}
\newcommand{\qwek}{\mathtt{k}}
\newcommand{\qwex}{\mathtt{x}}
\newcommand\qwe{\qwes,\qwek,\qwex}

The initial state is $\sysc{\ \mathtt{prot,\ read}\mid \mathtt{prot,\ read}\vdash \emptyset}$.
We denote the body of \texttt{prot} by $E$. The
transition corresponding to the secret being leaked is shown in Fig.~\ref{fig:leak}.
\begin{figure}[t]\small
\[
\begin{aligned}
    &\sysc{ N_0\mid P_0\vdash \emptyset}
    \xrightarrow[\emptyset]{\mathtt{call}\ \mathtt{prot}\, (),k}
    \langle N_0, k\mid P_0, k\vdash
    \emptyset,-,E,{k}\rangle\\
    \xrightarrow[\qquad\qquad]{\qquad}^*& \langle N_1, k, a_0,a_1 \mid P_0, k\vdash
    (\qwes\mapsto a_0,p\mapsto a_1, \qwex\mapsto 0),\\
    &\quad(\square;\mathtt{ if(*\mathit{x}==*\mathit{p})\ then\
      *\mathit{s}\ else\
      *\mathit{p}}) \circ(\mathtt{x=}\square)\circ(\mathtt{read}(\square)),() ,{k}\rangle\\
    \xrightarrow[\emptyset]{\mathtt{call\ read}\, (), k'}
    &\langle\!\langle N_1, k, k',a_0, a_1 \mid P_1 \vdash (\qwes\mapsto a_0,\qwek\mapsto a_1, \qwex\mapsto
    0, k'\mapsto (\f,k))\rangle\!\rangle\\
    \xrightarrow[\emptyset]{\quad\mathtt{ret}\ a_2,k'}& \langle
    N_2 \mid P_1,a_2\vdash(\qwes\mapsto a_0,\qwek\mapsto a_1, \qwex\mapsto 0,k'\mapsto (\f,k)),\f,a_2,k\rangle \\
    \xrightarrow[\qquad\qquad]{\qquad}^*&\langle N_2\mid P_1,a_2\vdash (\qwes\mapsto a_0,\qwek\mapsto a_1, \qwex\mapsto a_2,k'\mapsto (\f,k)), -, a_1, k\rangle\\
    \xrightarrow[\emptyset]{\quad\mathtt{ret}\ a_1,k}&
    \sysc{N_2 \mid P_2,a_2,a_1\vdash  (\qwes\mapsto a_0,\qwek\mapsto a_1, \qwex\mapsto a_2,k'\mapsto (\f,k))}\\
    \xrightarrow[\emptyset]{\quad\mathtt{ret}\ a_1,k'}& \langle
    N_2 \mid P_1,a_2,a_1\vdash (\qwes\mapsto a_0,\qwek\mapsto a_1, \qwex\mapsto a_2,k'\mapsto (\f,k)), \f,a_1,k\rangle \\
    \xrightarrow[\qquad\qquad]{\qquad}^*&\langle N_2 \mid P_1,a_2,a_1\vdash (\qwes\mapsto a_0,\qwek\mapsto a_1, \qwex\mapsto a_1,k'\mapsto (\f,k)), -, a_0, k\rangle\\
    \xrightarrow[\emptyset]{\quad\mathtt{ret}\ a_0,k}&
    \sysc{N_2 \mid P_2,a_2,a_1,{a_0}\vdash  (\qwes\mapsto a_0,\qwek\mapsto a_1, \qwex\mapsto a_2,k'\mapsto
      (\f,k))}.
  \end{aligned}
\]
  Above, $\f=(\square;\mathtt{ if(*\mathit{x}==*\mathit{k})\ then\ *\mathit{s}\ else\ *\mathit{k}})\circ(\mathtt{x=}\square)$, $N_0=P_0=\{\mathtt{prot, read}\}$, $N_1=N_0\cup\{\qwe\}$, $N_2=N_1\cup\{k,k',a_0,a_1,a_2\}$ and $P_1=P_0\cup\{k,k'\}$.
  \caption{Secret $a_0$ leaks.}
\label{fig:leak}
\end{figure}
The labelled transitions are the interactions between the program
and the system and are interpreted as follows:
\begin{compactenum}
\item system calls \texttt{prot()} giving continuation $k$ 
\item program calls \texttt{read()} giving fresh continuation $k'$
\item system returns (from \texttt{read}) using $k'$ and producing
  fresh name~$a_2$
\item program returns (from \texttt{prot}) leaking local name $a_1$
  stored in $\mathtt{k}$
\item system uses $k'$ to fake a second return from \texttt{read}, using
  the just-learned name $a_1$ as a return value \label{attack}
\item with $a_1$ the program now returns the secret $a_0$ stored in
  $s$ to the environment.
\end{compactenum}
Values of $a_2$ are omitted as they do not affect the transitions.

The critical step is~(\ref{attack}), where the system is using a
continuation in a presumably illegal, or at least unexpected, way.
This attack could be executed in a language with \texttt{call-cc}-like
control features, but these are lacking from our language.  We do not even need a
richer ambient language to show how a \emph{system-level} attack can
be actually implemented. Surprisingly, all we need is an
implementation of \texttt{read()} which will wait to receive a value
from the attacker, and a \texttt{main()} function which calls
\texttt{prot()} and reports the value.

We execute the attack by running the (closed) program in a
\emph{virtual machine} in the following way:
\begin{enumerate}
\item execute the program normally until the \texttt{read} function is
  called;
\item pause the virtual machine, save its state and exit;
\item duplicate the file storing the state of the virtual machine and
  re-start one instance of the virtual machine;
\item feed an arbitrary value to \texttt{read()};
\item when the program terminates normally remember the final value,
  which corresponds to $a_1$, stored in \texttt{k};
\item re-start the other instance of the virtual machine;
\item feed $a_1$ to \texttt{read()};
\item when this instance of the program terminates normally it leaks
  the secret $a_0$ from \texttt{s}.
\end{enumerate}
Note that the interaction between the attacker and the two instances
of the program correspond precisely to the labelled actions in the
attack.

What is remarkable about this attack is that both the term and the
context are written in a simple programming language that cannot
implement the attack! The attack happens because of a system-level
action, the cloning of a virtual machine. Also note that this is not a
theoretical attack. Our language is a subset of C and it can be
compiled, with small syntactic adjustments, by conventional C
compilers and executed on conventional operating systems. Any
virtualisation platform such as VMWare or VirtualBox can be used to
express this attack.

\subsection{Equivalence}\label{sec:eq}

Functional Compositionality gives an internal
consistency check for the semantics. This already shows that our
language is ``well behaved'' from a system-level point of view.  In
this section we want to further emphasise this point. We can do that by
proving that there are nontrivial equivalences which hold. There are
many such equivalences we can show, but we will choose a simple but
important one, because it embodies a principle of locality for
 state.

This deceptively simple example was first given
in~\cite{MeyerS88} and establishes the fact that a \emph{local}
variable cannot be interfered with by a \emph{non-local}
function. This was an interesting example because it highlighted a
significant shortcoming of \emph{global state} models of imperative
programming. 
Although not
pointed out at the time, functor-category models of state developed
roughly at the same time gave a mathematically clean solution for this
equivalence, which followed directly from the type structure of the
programming language~\cite{Tennent90}.

For comparing SLS LTSs we can use a simpler notion of bisimulation which relates configurations, and modules, that have common public names.
\begin{definition}
$\mathcal R$ is a simulation if, whenever $(X_1,X_2) \in\mathcal R$, 
\begin{compactitem}
\item $X_1$ and $X_2$ have the same public names;
\item $X_1\rightarrow X_1'$ implies $(X_1',X_2) \in\mathcal R$;
\item $X_1\redd{\alpha}{s}X_1'$ implies $(\pi\cdot X_2)\redd{\alpha}{s}X_2'$ and $(X_1',X_2') \in \mathcal R$, for some name permutation $\pi$ such that $\pi(a)=a$ for all public names $a$ of $X_1$ and $X_2$.
\end{compactitem}
$\mathcal R$ is a bisimulation if it and its inverse are simulations. We say that modules $M_1$ and $M_2$ are bisimilar if there is a bisimulation $\mathcal R$ such that $(S_{M_1}^0,S_{M_2}^0)\in\mathcal{R}$.
\end{definition}
\begin{proposition}\label{prop:cong}
Bisimulation is a congruence for module composition $- \cdot -$. 
\end{proposition}
The proof uses the reduction of syntactic composition to semantic composition then uses
Prop.~\ref{prop:compsls} to show that bisimulation is preserved by semantic
composition with the same module, which is immediate. This is unsurprising, since system-level
bisimilarity is more fine-grained than contextual equivalence in the programming language. 

It is straightforward to check that the following three programs have
bisimilar SLS transition systems:
\begin{verbatim}
  export f; import g; decl f() {local x; g(); return *x;}
  export f; import g; decl x; decl f() {g(); return *x;}
  export f; import g; decl f() {g(); return 0;}
\end{verbatim}
Intuitively, the reason is that in the first two programs
\texttt{f}-local (module-local, respectively) variable \texttt{x} is
never visible to non-local function \texttt{g}, and will keep its
initial value, which it 0.
The bisimulation relation is
straightforward as the three LTSs are equal modulo silent transitions
and permutation of private names for~\texttt x. 
Other equivalences, for example in the style of \emph{parametricity}~\cite{OHearnT95}
also hold, with simple proofs of equivalence via bisimulation.

\section{Conclusion}

The Dolev-Yao-like characterisation of the Opponent in this semantics
suggests that this is a model suitable for modelling security properties. The
system-level semantics presupposes certain strong guarantees of
secrecy and integrity for the combined execution environment
consisting of compiler and operating system: certain location names
must be kept secret; the Program source code cannot be altered; even
if the name of a function or continuation is disclosed the names used
in the function and in the continuation remain secret. All these
requirements can be gathered under the principle that \emph{the  System
can make no low-level attacks against the Program.} This is also
consistent with the Dolev-Yao principle that the attacker can
manipulate messages, but without breaking cryptography.

Compilers such as \texttt{gcc} do not implement a
system-level semantics since locations are not secret and the code
layout is known, allowing low-level attacks. Most
security violation of C code are through low-level attacks such as
buffer overflows. However, there are significant research
and industrial efforts to produce tamper-proof code through
techniques such as address layout
randomisation~\cite{JagadeesanPRR11,ShachamPPGMB04}, address
obfuscation~\cite{Bhatkar03}, instruction-set
obfuscation~\cite{GauravKP03} or secure processors~\cite{Suh03}. A
system-level semantics gives a semantically-directed specification for
a tamper-proof compiler.
System-level semantics  also gives a
 basis for the study of security properties of programs
compiled with such tamper-proof compilers, highlighting
\emph{logical} attacks, such as the secrecy violation in
Sec.~\ref{sec:vio}. The system does not
\emph{guess} any of the secrets of the program and it does not tamper
with its code, but it clones it wholesale then plays the two instances
against each other, a typical replay attack. Conversely, the equivalences
of Sec.~\ref{sec:eq} present opportunities for optimisations which hold
not only relative to certain compilers, but to a more comprehensive
concept of execution environment.

\bibliographystyle{abbrv}

\newpage
\appendix

\section{Nominal Sets}\label{App:NS}
It is handy to introduce here some basic notions from the theory of nominal sets~\cite{GabbayP99}.
We call \emph{nominal structure} any structure which may contain names, i.e.~elements of $\NAM$, and we denote by $\mathit{Perm}$ the set of finite permutations on $\NAM$ which are sort-preserving (i.e.~if $a\in\NAM_\lambda$ then $\pi(a)\in\NAM_\lambda$, etc.). For example, $\mathtt{id}=\{(a,a)\,|\,a\in\NAM\}\in\mathit{Perm}$. For each set $X$ of nominal structures of interest, we define a function
$ \_\cdot\_\ : \mathit{Perm}\times X \rightarrow X $
such that $\pi\cdot(\pi'\cdot x)=(\pi\circ\pi')\cdot x$\ and\ $\mathtt{id}\cdot x=x$, for all $x\in X$ and $\pi,\pi'\in\mathit{Perm}$. $X$ is called a \emph{nominal set} if all its elements involve finitely many names, that is, for all $x\in X$ there is a finite set $S\subseteq\NAM$ such that $\pi\cdot x=x$ whenever $\forall a\in S.\pi(a)=a$. 
The minimal such set $S$ is called the \emph{support} of $x$ and denoted by $\nu(x)$.
For example, $\NAM$ is a nominal set with action $\pi\cdot a = \pi(a)$, and so is $\mathcal{P}_{\mathsf{fn}}(\NAM)$ with action \
$\pi\cdot S = \{\pi(a)\ |\ a\in S\}$.

Also, any set of non-nominal structures is a nominal set with trivial action $\pi\cdot x=x$.
More interestingly, if $X,Y$ are nominal sets then so is $X\times Y$ with action \
$\pi\cdot (x,y)= (\pi\cdot x,\pi\cdot y)$. This extends to arbitrary products and to strings.
Also, if $X$ is a nominal set then so is the set $\bigcup\nolimits_{n\in\omega}(\{1,...,n\}\rightharpoonup X)$ with action \
$\pi\cdot f=\{(i,\pi\cdot x)\ |\ (i,x)\in f\}$. Finally, if $X,Y$ are nominal sets then so is the set $X\rightharpoonup_{\mathsf{fn}}Y$
with action \
$\pi\cdot f=\{(\pi\cdot x,\pi\cdot y)\ |\ (x,y)\in f\}$. 

\section{Proof of Proposition~\ref{prop:compsls}}\label{app:proofcompsls}

\begin{proof}
\newcommand\qwebar[1]{#1_{\mathsf{i}}}
For 1, let $X_1=\proc{N_1\mid P_1\vdash s_1,\f\circ f(\square),v,k}$, $X_2=\sysc{N_2\mid P_2\vdash s_2}$ and the $\tau$-transition being due to an internal transition with label $(\qwebar{s},\call f,v,k')$. Thus, 
$X_1'=\sysc{N_1'\mid P_1'\vdash s_1'}$, $X_2'=\proc{N_2'\mid P_2'\vdash s_2',f(\square),v,k'}$, and so $R(X_1\comp{\pub}{s}X_2)=\proc{N_0\mid \pub\vdash s_0,\f_0,v,k_0}$ and 
$R(X_1'\comp{\pub}{s'}X_2')=\proc{N_0'\mid \pub\vdash s_0',\f_0',v,k_0'}$. Computing $K,K'$ as above, we have $K'=K\cup\{k'\}$. 
Moreover, $s_1'=s_1[k'\mapsto(\f,k)]$ and $s_2'=s\cup(s_2\setminus P_2)$, 
so $(\f_0',k_0')=(s_1',s_2')_{K'}(f(\square),k')=(s_1',s_2')_{K'}(\f\circ f(\square),k)$. Since $k'$ is fresh, $(s_1',s_2')_{K'}(\f\circ f(\square),k)=(s_1,s_2)_{K}(\f\circ f(\square),k)=(\f_0,k_0)$. Moreover, $N_0=(N_1\cup N_2)\setminus K$ and $N_0'=(N_1'\cup N_2')\setminus K'=(N_1\cup\{k'\}\cup N_2\cup \nu(v,\qwebar s))\setminus K'$. As $\nu(v,\qwebar s)\subseteq N_1$ and $k'\in K'$, we get $N_0=N_0'$.
Finally, $s_0=\hat s_2[\hat s_1]\setminus K$ and $s_0'=\hat s_1'[\hat s_2']\setminus K'$. 
Thus, $s_0'=\hat s_1[\hat s_2']\setminus K=\hat s_1[s'\cup (\hat s_2\setminus\loc{P_2})]\setminus K$. Moreover, $s'=s_1\upharpoonright\loc{P_1'}$ so $s_0'=\hat s_1[\hat s_2\setminus\loc{P_2}]\setminus K$. But now note that $\dom(s_2\setminus\loc{P_2})\cap\dom(s_1)=\emptyset$: by the previous lemma, $\dom(s_1)$ and $\dom(s_2)$ share no continuation names, and if $a$ is a location name in $\dom(s_2)\setminus P_2$ then $a\notin N_1$. Thus, $s_0=s_0'$. Similarly if the $\tau$-transition is due to an internal return.  
 
Item 2 is straightforward. For 3, the only interesting issue is establishing that if $X_1\comp{\pub}{s}X_2$ is in such a form that a $\tau$-transition needs to take place then the latter is possible.
This follows directly from the definition of the transitions and the conditions of the previous lemma. In the following cases we consider call transitions; cases with return transitions are treated in a similar manner.

For 4, let $X_1=\sysc{N_1\mid P_1\vdash s_1}$, $X_2=\sysc{N_2\mid P_2\vdash s_2}$, $\alpha=(s',\call f,v,k)$ and suppose the transition is due to $X_1$ reducing to $X_1'=\proc{N_1'\mid P_1'\vdash s_1',f(\square),v,k}$ with label $(\qwebar{s},\call f,v,k)$. We have $\pub'=\pub\cup\nu(v,k,\qwebar{s}\setminus\pr)$, $\pr=(N_1\cup N_2)\setminus\pub$, $s'=\qwebar{s}\upharpoonright\pub'$ and 
$\nu(v,\qwebar{s}\setminus\pr)\cap\pr=\emptyset$.
Let
$R(X_1\comp{\pub}{s}X_2)=\sysc{N_0\mid \pub\vdash s_0}$. 
As $k\notin\dom(s_1)$ and $k\notin\nu(X_2)\setminus\pub_S$, by previous lemma we obtain $k\notin\dom(s_0)$, so
the latter reduces to $\proc{N_0'\mid P\vdash s_0',f(\square),v,k}$ with transition $(s''',\call f,v,k)$, for any appropriate $s'''$. In fact, if $\nu(v,s')\cap N_0\subseteq\pub$ then we can choose $s'''=s'$. 
Indeed, $(\nu(v,s')\cap N_0)\setminus\pub\subseteq\nu(v,s')\cap (N_0\setminus\pub)\subseteq \nu(v,s')\cap\pr=\nu(v,\qwebar{s}\upharpoonright\pub')\cap\pr=\nu(v,\qwebar{s}\setminus\pr)\cap\pr=\emptyset$. 
Let $R(X_1'\comp{\pub'}{s''}X_2)=\sysc{N_0''\mid \pub'\vdash s_0''}$. We can see that $N_0'=N_0''$. 
Also, $P=\pub\cup\{k\}\cup\nu(v,s')$ while $\pub'=\pub\cup\nu(v,k,\qwebar{s}\setminus\pr)=\pub\cup\nu(v,k,s')$. Moreover, $s_0'=s'\cup(s_0\setminus\loc{\pub})=s'\cup((\hat s_1[s_{12}]\cup \hat s_2[s_{12}])\setminus (K\cup\loc{\pub}))$ with $s_{12}=s\upharpoonright(P_1\cap P_2)$, and $s_0''=\hat{s}_2[\hat{s}_1']\setminus K'=\hat{s}_2[\qwebar{s}\cup(\hat{s}_1\setminus\loc{P_1})]\setminus K'$. Note that $K'=K$.
Moreover, $s_0'$ and $s_0''$ agree on the domain of $s'$ and on continuation names. Also, if location name $a\in N_0'\setminus N_0$ then $a\in\nu(v,s')$ and thus $a\in\dom(s')$. Thus, we need to show that $s_0',s_0''$ agree on location names $a$ from $N_0\setminus\pub$. 
If
$a\in N_1\setminus P_1$ then $s_0'(a)=s_1(a)=s_0''(a)$, and similarly if in $N_2\setminus P_2$ using the fact that $(N_2\setminus P_2)\cap N_1=\emptyset$. Finally, if $a\in P_1\setminus\pub=P_2\setminus\pub$ then $s_0'(a)=s(a)=\qwebar{s}(a)=s_0''(a)$, by restrictions on $\qwebar{s}$.

Now let $X_1=\proc{N_1\mid P_1\vdash s_1,\f\circ f(\square),v,k}$, $X_2=\sysc{N_2\mid P_2\vdash s_2}$, $\alpha=\call f,v,k'$ and suppose the transition is due to $X_1$ reducing to $X_1'=\sysc{N_1'\mid P_1'\vdash s_1'}$ with label $(\qwebar{s},\call f,v,k')$. 
We have $(\pub',s'')=(\pub,s)[v,\qwebar s]$ and $s'=s''\upharpoonright \pub'$.
We can assume, by definition, that  $(s_1,s_2)_K(\f\circ f(\square),k)=(\f_0\circ f(\square),k_0)$,
so $R(X_1\comp{\pub}{s}X_2)=\proc{N_0\mid \pub\vdash s_0,\f_0\circ f(\square),v,k_0}$. As $f$ is not defined in either of the modules and $k'$ is completely fresh, the latter reduces to $\sysc{N_0'\mid P\vdash s_0'}$ with transition $(s''',\call f,v,k')$. 
Let $R(X_1'\comp{\pub'}{s''}X_2)=\sysc{N_0''\mid \pub'\vdash s_0''}$. 
It is easy to see that $N_0'=N_0''$. Moreover, $s_0'=s_0[k'\mapsto (\f_0,k_0)]$ and $s_0''=(\hat{s}_1'[s''_{12}]\cup \hat s_2[s''_{12}])\setminus K$ where $s''_{12}=s''\upharpoonright(P_1'\cap P_2)$. Note that $K'=\cont{P_1'\cap P_2}=K$ and $s_0''(k')=\hat{s}_1'(k')=(s_1',s_2)_{K'}(\f,k)=(\f_0,k_0)$. Also, $s_0'$ and $s_0''$ agree on all other continuation names. Thus, in order to establish that $s_0'=s_0''$, it suffices to show that $s_2[s_1]$ and ${s}_1[s''_{12}]\cup s_2[s''_{12}]$ agree on locations. 
From the previous lemma, $s''$ agrees with $s_1$ on locations in $P_1'$ and with $s_2$ on locations in $P_2\setminus P_1'$, and so $s''\subseteq s_2[s_1]$. Thus, $\loc{{s}_1[s''_{12}]\cup s_2[s''_{12}]}=\loc{s_1\cup (s_2\setminus P_1')}=\loc{s_2[s_1]}$.

For public names, we have $P=\clos{s_0}{\pub\cup\nu(v)}\cup\{k'\}=\clos{s_0'}{\pub\cup\nu(v,k')}$ while $\pub'=\clos{s''}{\pub\cup\nu(v,k')}$. As $\cont{P}=\cont{\pub}\cup\{k'\}=\cont{\pub'}$, we can focus on location names. We have $s''\subseteq s_0'$ and, moreover, $\dom(s'')=\loc{P_1'\cup P_2}\supseteq\loc{\pub\cup\nu(v,k')}$, thus $P=\pub'$.
Finally, $s'=s'''$ follows from the fact that these are restrictions of the final stores to the final sets of public location names.

For 5, let $X_1=\sysc{N_1\mid P_1\vdash s_1}$, $X_2=\sysc{N_2\mid P_2\vdash s_2}$, $R(X_1\comp{\pub}{s}X_2)=\sysc{N_0\mid \pub\vdash s_0}$ and $\alpha=(s',\call f,v,k)$. We have that $f$ is defined in $M\cdot M'$ so WLOG assume that it is defined in $M$. Then, $X_1$ reduces to $X_1'=\proc{N_1'\mid P_1'\vdash s_1',f(\square),v,k}$ with $(\qwebar{s},\call f,v,k)$, $\qwebar{s}=s'\cup(s\setminus\pub)$, if the relevant conditions for S-P calls are satisfied. 

If $k\in\dom(s_1)$ then, by lemma, $k\notin\pub_S$. By assumption, $k\in\pub$ so $k\notin\cont{P_1\cup P_2}\setminus\pub$ and thus, by lemma, $k\notin\cont{P_1\cap P_2}\setminus\pub_S$ so $k\notin P_2$. But the latter would imply $k\in\dom(s_0)$, which is disallowed by definition. Thus, $k\notin\dom(s_1)$. 

Moreover, if $a\in\nu(v,\qwebar{s})\cap(N_1\setminus P_1)=\nu(v,s')\cap(N_1\setminus P_1)$ then $a\in\nu(v,s')\cap(N_0\setminus P_1)$ and $a\notin P_2$, so $a\in\nu(v,s')\cap(N_0\setminus(P_1\cup P_2))\subseteq \nu(v,s')\cap(N_0\setminus\pub)$, thus contradicting the conditions for the transition $\alpha$. We still need to check that $s_1\upharpoonright \loc{P_1}\sqsubseteq\qwebar{s}=s'\cup(s\setminus\pub)$. Given that $s_0\upharpoonright\loc{\pub}=(s_1[s_{12}]\cup s_2[s_{12}])\upharpoonright\loc{\pub}\sqsubseteq s'$, the condition follows from the previous lemma. We therefore obtain a transition from $X_1\comp{\pub}{s}X_2$ to $X_1'\comp{\pub'}{s''}X_2$; the relevant side-conditions are shown to be satisfied similarly as above.
Finally, working as in 4, we obtain $R(X_1'\comp{\pub'}{s''}X_2)=Y$ and $s'=s'''$. 

Now let $X_1=\proc{N_1\mid P_1\vdash s_1,\f\circ f(\square),v,k}$, $X_2=\sysc{N_2\mid P_2\vdash s_2}$, 
$R(X_1\comp{\pub}{s}X_2)=\proc{N_0\mid \pub\vdash s_0,\f_0\circ f(\square),v,k_0}$ and 
$\alpha=\call f,v,k'$. By hypothesis, $k'$ is fresh and therefore $X_1$ reduces to $X_1'=\sysc{N_1'\mid P_1'\vdash s_1'}$ with $(\qwebar{s},\call f,v,k')$, and  thus $X_1\comp{\pub}{s}X_2$ reduces to $X_1'\comp{\pub'}{s''}X_2$ with $(s''',\call f,v,k')$. Working as in 4, $R(X_1'\comp{\pub'}{s''}X_2)=Y$ and $s'=s'''$.
\end{proof}

\end{document}